\documentclass[11pt,twoside]{article}
\usepackage{asp2004}
\usepackage{psfig}
\usepackage{epsf}
\usepackage{graphics}
\usepackage{lscape}
\def\edcomment#1{\iffalse\marginpar{\raggedright\sl#1\/}\else\relax\fi}
\reversemarginpar

\begin{document}
\title{The Initial Mass Function of Stars}
 \author{Bruce G. Elmegreen}
\affil{IBM T.J. Watson Research Center, 1101 Kitchawan Road,
Yorktown Hts., NY 10598 USA, bge@watson.ibm.com}

\begin{abstract}
Observations and theory of the IMF are briefly reviewed. Slight
variations have been observed, although they are difficult to
prove unambiguously. Most detailed theoretical models reproduce
the IMF, but because they use different assumptions and
conditions, there is no real convergence of explanations yet.
\end{abstract}

\vspace{-0.5cm}
\section{Introduction}
Stars are born with a mass distribution function that has a nearly
constant stellar mass in each log interval of mass above about
$0.5$ M$_\odot$. A common fit to observations is a power law IMF
with a slope $\alpha$ between $-2.1$ and $-2.6$, i.e., for equal
intervals of mass, $dn(M)=n_0 M^{-\alpha}dM$. The center of this
range is $\alpha=-2.35$, which is the Salpeter (1955) IMF. The
coefficient $n_0$ is given by the total stellar mass or the
cluster mass, $n_0=\left(\alpha-2\right)M_{min}^{\alpha-2}M_{cl}$,
where the minimum star mass in the power law part is $M_{min}$ and
the cluster mass in this expression is assumed for simplicity to
be the mass of the power law part. At low mass, the IMF rises less
steeply than the Salpeter function.  For an IMF with a typical
slope of $-1.35$ between 0.05 and 0.5 M$_\odot$ and a slope of
$-2.35$ between 0.5 and 50 M$_\odot$, the fractional mass in each
factor of ten interval of mass, from $0.05$ to 50 M$_\odot$, is
21\%, 54\%, and 24\%.  The peak between 0.5 and 5 M$_\odot$ is a
characteristic mass.  It may be defined more precisely as the mass
where the steep and shallow power laws meet.  In the Kroupa (2001)
IMF, it is 0.5 M$_\odot$.

The IMF is not constant from region to region (Scalo 1998), but
many of the observed variations in slope and characteristic mass
could be stochastic for the typically small samples that are
observed (Elmegreen 1999; Kroupa 2001).  Thus the observations are
consistent with typical star clusters sharing a common probability
distribution function for the IMF, with more or less random
sampling around that function as stars form.

Some IMF variations seem more significant than this because they
recur for regions with peculiar physical properties, such as
starbursts. Starbursts have long been suspected of having a
flatter slope at intermediate to high mass (Rieke et al. 1980,
1993). However, the early inferences of top-heavy IMFs for
starbursts appear now to be premature because of excessive
extinction corrections and other problems (Devereux 1989; Satyapal
1995; see review in Elmegreen 2005). The most recent reports of
IMF flattening, which are for super star clusters (see below),
have also been questioned lately, as there could be a lack of
dynamical equilibrium in these clusters (Bastian \& Goodwin 2006)
or non-isothermal velocity dispersions.  IMF flattening has also
been reported for the young bursting phases of elliptical galaxies
and galaxy clusters (see references below). Conversely, an IMF
steepening has been claimed for low surface brightness disks (Lee
et al. 2004). If these subtle variations are real, then they
suggest a slight trend toward more massive stars with increasing
density (Elmegreen 2004).

Most systematic variations in the IMF are small, just several
tenths in $\alpha$ (except for the LMC ``extreme'' field, see
below). This has led to the idea of a universal IMF. The reason
for such general uniformity is not clear. It could be that
hierarchical fragmentation of any type gives the basic $dn/dM\sim
M^{-2}$ shape (equal mass in equal log intervals of mass for a
pure hierarchy), with a slight preference for lower masses giving
the Salpeter and steeper functions. Such a preference could come
from slightly faster collapse at low mass, which seems natural for
a hierarchical medium as the density is usually larger at lower
mass (Elmegreen 1999). The basic IMF form could also come from
accretion (Bonnell et al. 2001) or coalescence (Shadmehri 2004) or
a combination of these processes (Bonnell et al. 2006). On the
other hand, the intrinsic IMF may not be a simple power law -- it
could be a composite of functions reflecting different processes
in different mass ranges (Elmegreen 2004).

The characteristic mass for star formation is not well understood,
nor are its variations and trends. It is commonly thought to be
the thermal Jeans mass in the cloud core, but this is a poorly
defined quantity when the core pressure should vary with position.
For example, it is not clear why the most massive dense clusters,
which appear to have high internal pressures, have the same
characteristic mass in the IMF as star formation in lower pressure
regions. Is there thermal feedback involving both temperature and
pressure that keeps the Jeans mass relatively constant? Or does
the characteristic mass have a different origin? We don't know.

The IMF extends down from the characteristic mass by more than a
factor of $\sim30$ (Luhman et al. 2000; Lucas et al. 2005). Is the
star formation process completely different for these low masses
(e.g., Bate, Bonnell \& Bromm 2002)? Is the Jeans mass highly
reduced in some places (Padoan et al. 2004)?  Does the lower mass
limit for star formation vary from region to region?

\section{Observations}

\subsection{Clusters}

The IMF is observed in many types of regions with diverse
selection effects and limitations. Star clusters have the
advantage that all of the stars have about the same age and
distance, but mass segregation, field contamination, small number
statistics, and evaporation can be problems. In OB associations
too, the stars have about the same distance, but there is usually
a range of ages and the possibility of dispersal over time into
the surrounding field. Field IMFs can have a large number of
stars, making the statistical accuracy large, but the inferred
mass function, which is expressed per unit volume or area, depends
on the star formation history, vertical disk heating, selective
drift away from clusters, and other things. The IMFs in whole
galaxies have been determined from abundance ratios (e.g., Fe
comes from low and high mass stars whereas O comes from high mass
stars) and star counts, but poor resolution, faintness, and
unknown star formation histories can be problems. Extinction
variations can affect all of these measurements.

Generally only massive clusters have been used to determine the
high mass IMF. Low mass clusters do not usually have high mass
stars. Massive stars also form in peripheral gas, however, because
of triggering or independent processes. It is not clear how to
include such regions in the IMF. Should their stars add to the
total in the cluster or stand alone? The answer may depend on
whether one thinks the cluster environment is important for the
IMF.

Many dense clusters are observed to have about the Salpeter IMF
slope at intermediate to high mass: R136 in 30 Dor (Massey \&
Hunter 1998), h and $\chi$ Persei (Slesnick, Hillenbrand \& Massey
2002), NGC 604 in M33 (Gonzalez Delgado \& Perez 2000), NGC 1960
and NGC 2194 (Sanner et al. 2000), NGC 6611 (Belikov et al. 2000),
and many others. The upper Scorpius OB association has a steeper
IMF, $dn/dM\propto M^{-2.8}$ between 0.6 and 2 M$_\odot$ and
$dn/dM\propto M^{-2.6}$ between 2 and 20 M$_\odot$ (Preibisch et
al. 2003). The intermediate stars in W51 have a steep slope too,
$M^{-2.8}$, but at high mass there is a clear excess of stars
compared to this slope in two of four stellar subgroups (Okumura
et al. 2000).  NGC 604, mentioned above, is particularly
interesting because there appear to be no clusters there, leading
Hunter et al. (1996) to suggest that the IMF is independent of
density.

Mass segregation is a severe effect in most clusters. The slope of
the IMF is generally shallower in a cluster core than around the
periphery, which means that massive stars prefer the center. For
example, Sung \& Bessell (2004) found segregation in the massive
young cluster NGC 3606.  The Arches cluster in the Galactic center
has a shallow IMF (Yang et al. 2002; Stolte et al. 2005), but it
is not known whether these massive stars are all there ever was in
the cluster.  An envelope of low density stars could have been
tidally stripped (Kim et al. 2000; de Marchi, Pulone, \& Paresce
2006).

Some super star clusters have low mass-to-light ratios suggesting
a relative excess of high mass stars compared to a standard IMF.
Sternberg (1998) found a low M/L in NGC 1705-1, as if the slope is
shallower than $-2$ or the characteristic mass is higher than
normal. Smith \& Gallagher (2001) found a low M/L in M82F,
suggesting an inner cutoff of 2-3 M$_\odot$ for the Salpeter IMF.
Alonso-Herrero et al. (2001) observed a low M/L in the starburst
galaxy NGC 1614.  McCrady et al. (2003) found that in M82, MGG-11
is deficit in low mass stars.  Mengel et al. (2003) found the same
in NGC 4038/9.  Other super star cluster have normal IMFs: NGC
1569-A (Ho \& Filippenko 1996; Sternberg 1998), NGC 6946 (Larsen
et al. 2001), and M82, MGG-9 (McCrady et al. 2003).  The IMF is
difficult to measure in super star clusters. One has to observe
the velocity dispersion and radius of the cluster to get the
binding mass, and then combine this with the luminosity. The
velocity dispersion may vary with radius, however (e.g. NGC 6946),
and the value of radius is uncertain. The cluster could be
evaporating, out of equilibrium, non-isothermal, multi-component
or non-centralized (Bastian \& Goodwin 2006); the core could be
poorly resolved too. Field star corrections may be uncertain as
well.

The low mass part of the IMF varies from place to place also.
Studies by Luhman et al. (2000, 2003), Briceno et al. (2002),
Muench et al. (2003) and others suggest denser clusters have more
low mass stars and brown dwarfs.

\subsection{Field}

Scalo (1986) and Rana (1987) suggested that the slope of the IMF
in the local field is steeper than Salpeter, $-2.7$ to $-2.8$
(compared to $-2.35$). In an extensive study of the LMC field,
Parker et al. (1998) got the same slope, $-2.80\pm0.09$, for
masses larger than 2 M$_\odot$. Massey et al. (1995, 2002) got a
steeper slope, $-5$, in regions more than 30 pc from a Lucke \&
Hodge (1970) or Hodge (1986) OB association, complete down to 25
M$\odot$.  He assumed the star formation rate was constant over
the last 10 My and had 450 stars in the LMC sample. However, in a
field region near 30 Dor, the IMF slope was found to be more like
Salpeter, $-2.38\pm0.04$ for 7-40 M$_\odot$ (Selman \& Melnick
2005). For low mass field stars in the LMC, 0.6-1.1 M$\odot$,
Holtzman et al. (1997) got a slope of -2.0 to -3.1. In a field
region near the supershell LMC4, the slope is $-6$ for 0.9-2
M$_\odot$ and $-3.6$ for 0.9-6 M$_\odot$ (Gouliermis, Brandner \&
Henning 2005).  The origin of these variations is unknown. Field
regions near OB associations could be overpopulated with low mass
stars that disperse further from their birthplaces than high mass
stars (Elmegreen 1999; Hoopes, Walterbos \& Bothun 2001; Tremonti
et al. 2002).

Low Surface Brightness Galaxies have been suggested to have a
steep slope throughout, $-3.85$ for 0.1-60 M$_\odot$ (Lee et al.
2004). Red halos in BCD galaxies, and deep, stacked-image halos
around edge-on galaxy disks, also seem to have a steep IMF slope:
$-4.5$ (Zackrisson, et al. 2004).

The low mass IMF in the solar neighborhood has been observed by
many groups. Chabrier et al. (2005) suggested it continues to rise
at low mass, with no evidence for a turnover (on a log-log plot)
down to 0.12 M$_\odot$. Schultheis et al. (2006) observed the low
mass IMF in the thick disk of the Milky Way, where there seems to
be relatively few low mass stars compared to the low mass IMF in
the solar neighborhood.

\section{IMF Theory}

In the modern theory of star formation, turbulence causes dense
gas structures and fragmentation in intersecting shocks,
protostars form in the collapsing cores of these dense regions,
and then the protostars move around and accrete gas, possibly
coalescing or getting ejected from dense sub-clusters. Simulations
of these processes end up with a reasonable IMF (Bonnell et al.
2006).

Bate \& Bonnell (2005) tested the importance of the Jean mass with
an SPH simulation having no magnetic field. Two simulations
containing 50 Jeans-masses initially but with different values for
this Jeans mass showed that the mean mass of the fragments scaled
with the Jeans mass. Jappsen et al. (2005) also did SPH
simulations with no magnetic fields. They varied the equation of
state to give a ratio of specific heats less than 1 at low density
and greater than 1 at high density. The resulting IMF depended on
transition density: higher transition density resulted in a lower
Jeans mass and more cores. Still, the Salpeter IMF resulted.
Martel et al. (2005) did isothermal SPH with particle splitting
and no magnetic fields. They found that the core mass function
depends on resolution so that higher resolution gives a lower
characteristic mass. Tilley \& Pudritz (2005) used ZEUS-MP MHD
with a $256^3$ grid, considering different ratios of gravitational
to magnetic energy; bound cores with a reasonable IMF formed in
all of the highly supercritical cases. Padoan et al. (2004) did
adaptive mesh MHD in $1024^3$ cells and a Mach number of 6; they
formed brown dwarfs by turbulent fragmentation. Nakamura \& Li
(2005) did 2D MHD with magnetic diffusion enhanced by turbulent
compression. They found diffusion-regulated collapse in compressed
regions and a low SF efficiency.

\section{Conclusions}

IMF observations suggest a more or less constant IMF in many
diverse environments, although small but significant variations
have been found in all mass intervals. There are no generally
recognized systematic trends with these variations, however,
making it difficult to find causes. The only trend that has been
suggested is with density, in the sense that the IMF is sometimes
steep at intermediate to high mass at very low density, as in
field regions or low surface brightness galaxies, and it is
sometimes relatively shallow at very high densities, as in some
super star clusters or the bursting phases of galaxy formation.
Cloud fragmentation, protostellar coalescence, protostar
accretion, ejection from sub-clusters etc. could all play a role,
as demonstrated by simulations.

\end{document}